\documentclass[12pt]{scrartcl}

\usepackage{graphicx}
\usepackage{tabularx}
\usepackage{amsmath}

\newcommand{\lfo}{[(Li$_{1-x}$Fe$_x$)OH](Fe$_{1-y}$Li$_y$)Se}
\newcommand{\fea}{Fe$^{\rm a}$}
\newcommand{\feb}{Fe$^{\rm b}$}

\title{Coexistence of 3$d$-ferromagnetism\\ and superconductivity in \lfo}

\author{Ursula Pachmayr$^1$, Fabian Nitsche$^1$, Hubertus Luetkens$^2$,\\ Sirko Kamusella$^3$, Felix Br\"uckner$^3$, Rajib Sarkar$^3$,\\ Hans-Hennig Klauss$^3$, and Dirk Johrendt$^1$\footnote{Corresponding author, Email: johrendt@lmu.de}\\$^1$Department Chemie,
Ludwig-Maximilians-Universit\"at M\"unchen\\ Butenandtstr. 5-13 (Haus D), 81377 M\"unchen, Germany\\
$^2$Paul Scherrer Institut PSI, 5232 Villigen, Switzerland\\
$^3$Institut f\"ur Festk\"orperphysik, Technische Universit\"at Dresden,\\ 01062 Dresden, Germany
}

\date{}

\begin{document}

\maketitle

\begin{abstract}
Superconducting \lfo\ ($x \approx$ 0.2, $y \approx$ 0.08) was synthesized by hydrothermal methods and structurally characterized by single crystal X-ray diffraction. The crystal structure contains $anti$-PbO type (Fe$_{1-y}$Li$_y$)Se layers separated by layers of (Li$_{1-x}$Fe$_x$)OH. Electrical resistivity and magnetic susceptibility measurements reveal superconductivity at 43 K. An anomaly in the diamagnetic shielding indicates ferromagnetic ordering near 10 K while superconductivity is retained. The ferromagnetism emerges from the iron atoms in the (Li$_{1-x}$Fe$_x$)OH layer. Isothermal magnetization measurements confirm the superposition of ferromagnetic with superconducting hysteresis. The internal  ferromagnetic field is larger than the lower, but smaller than the upper critical field of the superconductor, which gives evidence for a spontaneous vortex phase where both orders coexist. $^{57}$Fe-M\"ossbauer spectra, $^7$Li-NMR spectra, and $\mu$SR experiments consistently support this rare situation, especially in a bulk material where magnetism emerges from a 3$d$-element.

\end{abstract}

\section*{Introduction}
Superconductivity expels magnetic flux from the interior of a solid, while  ferromagnetism generates it, thus both phenomena are antagonistic. Moreover, ferromagnetic order is usually detrimental to superconductivity because strong internal fields from aligned moments break Cooper pairs. Nevertheless, both phenomena are not mutually exclusive in all cases. After early investigations of alloys with magnetic rare earth atoms diluted in superconducting lanthanum metal \cite{Matthias-1958}, the first superconductors with spatially ordered arrays of magnetic atoms were the metallic molybdenum sulphides $RE$Mo$_6$S$_8$ ($RE$ = rare earth element), referred to as the Chevrel phases \cite{Chevrel-1971,Matthias-1972}. Among them, compounds with the strongly magnetic rare earth elements Tb-Er have superconducting critical temperatures ($T_{\rm c}$) around 2 K, and enter magnetically ordered states between 15 mK and 5 mK \cite{Ishikawa-1977,Lynn-1978}. A further example is ErRh$_4$B$_4$ where ferromagnetism destroys superconductivity at 1 K, while co-existence with antiferromagnetic ordering has been found in the borocarbides $RE$Ni$_2$B$_2$C \cite{Fertig-1977,Lynn-1997} and the ruthenate RuSr$_2$GdCu$_2$O$_8$ \cite{Lynn-2000}. Recently, the co-existence of superconductivity with ferromagnetic ordering of Eu$^{2+}$ ions in the iron arsenides EuFe$_2$(As$_{1-x}$P$_x$)$_2$ and Eu(Fe$_{1-x}$Ru$_x$)$_2$As$_2$  has been reported \cite{Cao-2011,Jiao-2012,Nandi-2014}. Such materials where the ferromagnetic ordering temperature $T_{\rm fm}$ is below  $T_{\rm c}$ are called ferromagnetic superconductors. Therein, both phenomena are usually spatially decoupled, and do not interact directly in the sense that the same electrons are responsible for both. The latter is discussed in superconducting ferromagnets with $T_{\rm fm} > T_{\rm c}$. Here the superconducting state emerges in a ferromagnetic metal (usually at mK temperatures), which gives evidence for exotic mechanisms like spin triplet pairing, for example in UGe$_2$ or URhGe which have been intensively studied \cite{Saxena-2000,Huxley-2001,Aoki-2001}.\bigskip

So far the extremely low temperatures, as well as the inertness of the rare earth 4$f$ shell hardly allowed chemical manipulation of these quite fascinating phenomena. This would be different if the ferromagnetic ordering emerges from $d$-elements, where the magnetic state is much more susceptible to the chemical environment. Materials where superconductivity coexists with 3$d$-ferromagnetism in a bulk phase are still unknown to our best knowledge \cite{Ruck-2011, Zhu-2012}. In this communication we report the synthesis, crystal structure and basic physical properties of the ferromagnetic superconductor \lfo, where magnetic ordering emerges from iron ions in the hydroxide layer at 10 K, which is sandwiched between iron selenide layers providing superconductivity up to 43 K.        

\section*{Results}

\subsection*{Synthesis and crystal structure}

\begin{figure}[h]
\centering
\includegraphics[width=13cm]{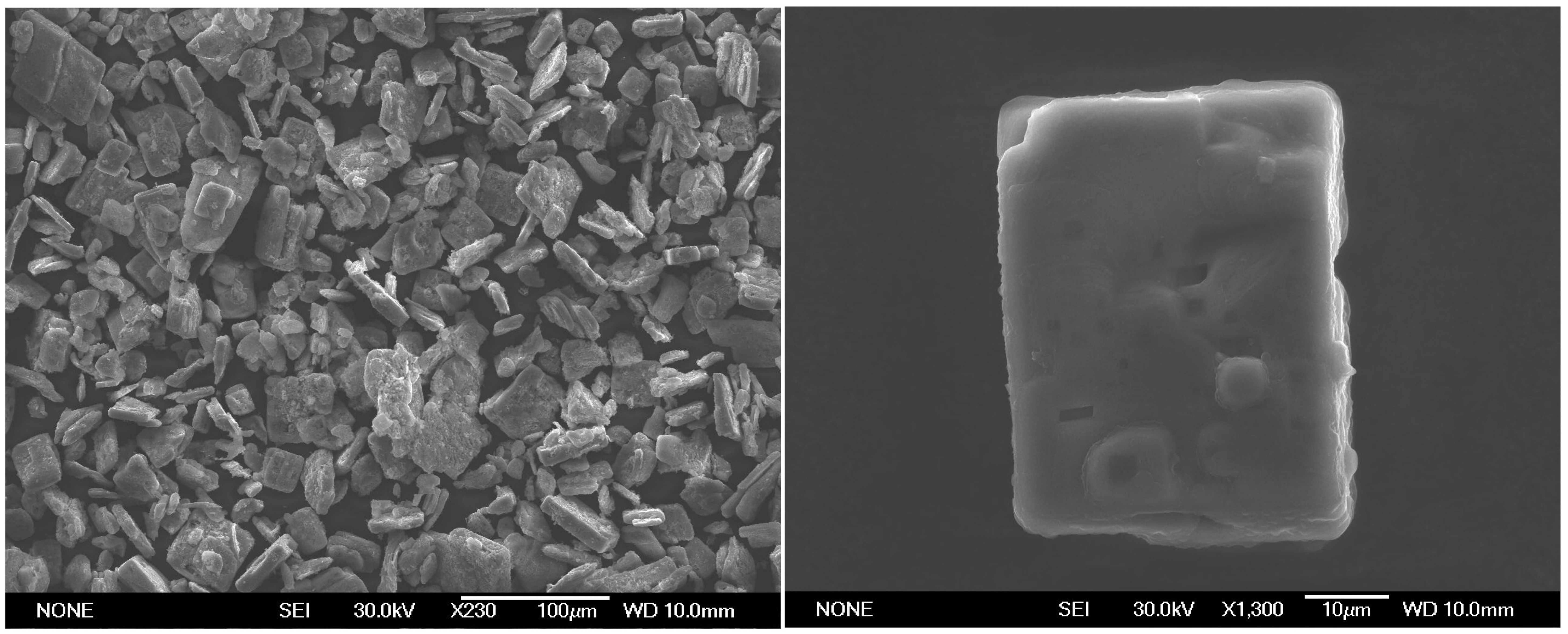}
\caption{\label{fig:sem}Left: SEM image of a \lfo\ sample; right: plate-like single crystal.} 
\end{figure}

\begin{figure}[h]
\centering
\includegraphics[width=12cm]{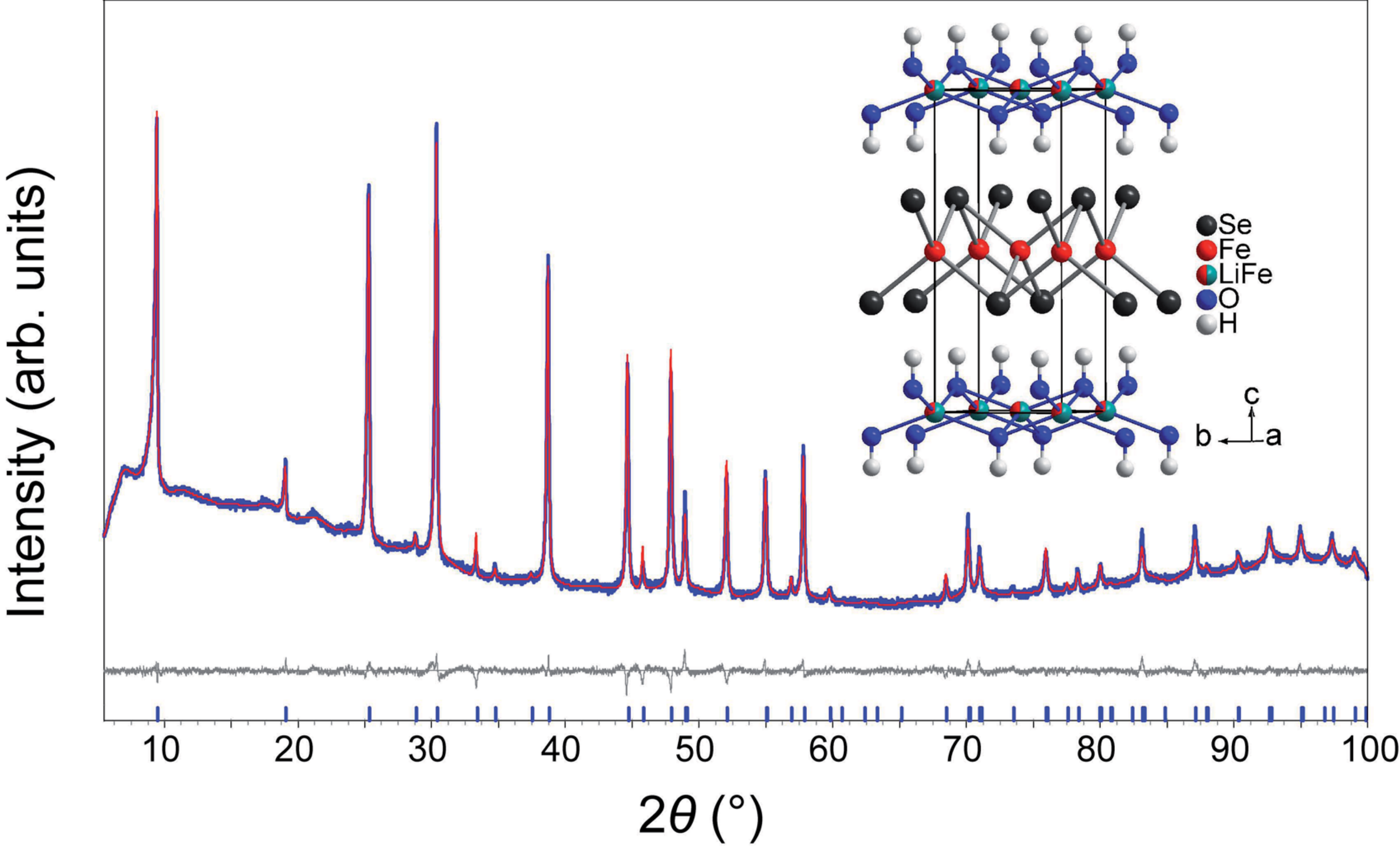}
\caption{\label{fig:xrd+struct}X-ray powder pattern (blue) with Rietveld-fit (red) and difference curve (grey). Insert: crystal structure of \lfo.} 
\end{figure}

Polycrystalline samples of \lfo\ were synthesized under hydrothermal conditions using a modified procedure given in \cite{Lu-2014}. Fig.~\ref{fig:sem} shows an electron microscope image of the sample and a typical plate-like crystal. A small specimen ($50 \times 40 \times 5~ \mu$m$^3$) was selected for the X-ray single crystal analysis. First structure  refinements using the data of LiFeO$_2$Fe$_2$Se$_2$ given by Lu \textit{et al.} \cite{Lu-2014} as starting parameters were not satisfying. A closer inspection revealed residual electron density at about 75 pm below the oxygen atoms which indicated additional hydrogen. Furthermore the $U_{33}$ component of the thermal displacement ellipsoid at the Fe/Li site was too large, which required a split position with Li shifted off the centre of the oxygen tetrahedra by 40 pm along the $c$-direction. Attempts to find an ordered model by twinning and/or symmetry reduction failed. Finally we detected a slight deficiency at the iron site in the FeSe layer. X-ray diffraction cannot distinguish between iron vacancies or a possible Fe/Li mixed site. Since Li-NMR shows two Li sites in the structure, we interpret the deficiency as Fe/Li mixing with $\approx$\,8\% Li. Using this model, the structure refinements rapidly converged to small residuals ($R$1 = 0.016). The crystallographic parameters are compiled in Table S1. In the following we denote iron in the hydroxide layer as Fe$^{\rm a}$ and in the FeSe layer as Fe$^{\rm b}$. By using these crystal data we were able to perform a Rietveld fit of the X-ray powder pattern, which revealed the identical structure and proved that the sample is virtually free from impurities within the sensitivity of laboratory X-ray powder diffraction. The crystal structure of \lfo\ is depicted in the insert of Fig.~\ref{fig:xrd+struct}. \textit{Anti}-PbO type layers of lithium-iron-hydroxide alternate with FeSe-layers. Unlike LiFeO$_2$Fe$_2$Se$_2$ \cite{Lu-2014} our compound is not an oxide but a hydroxide, where positively polarized hydrogen atoms point towards the negatively polarized selenium of the FeSe layer. The structure of the (Li$_{1-x}$Fe$_x$)OH layer is quite similar to LiOH itself, which likewise crystallizes in the $anti$-PbO-type \cite{Dachs-1959}.   

\bigskip

The Se-Fe-Se bond angles of the FeSe$_4$ tetrahedra are almost identical to those in binary $\beta$-FeSe  \cite{McQueen-2009}, while the Fe-Se bond lengths (241.4 pm) are slightly longer than in $\beta$-FeSe (239.5 pm). Thus no significant changes apply to the structure of the FeSe layer in \lfo, however, the tiny elongation of the Fe-Se bonds may already influence the electronic properties. The situation in the hydroxide-layer is more difficult. Iron is in a flattened tetrahedron of oxygen atoms with a Fe-O distance of 201.6 pm. This matches the sum of the ionic radii \cite{Shannon-1969} if iron is Fe$^{2+}$ (203 pm), but not if iron is Fe$^{3+}$ (189 pm). Thus we suggest Fe$^{2+}$ in the hydroxide layer, even though a tetrahedral coordination is rather unusual. Lithium in the centre of the flat oxygen tetrahedron would have Li-O distances of 201.6 pm, significantly longer compared to 196 pm in LiOH \cite{Mair-1978}. We suggest that this is the reason why Li is shifted along $c$, however, the Li position is not precise due to the low scattering power. This is even more the case for the hydrogen atom, where the refined O-H distance is 72(8) pm. Given the large error and the fact that X-H bond lengths from X-ray diffraction are usually underestimated by at least 10\,\%, we are not that far from the O-H distance in LiOH which was determined to 89 pm using neutron diffraction \cite{Dachs-1959}.       

The composition obtained from X-ray diffraction is (Li$_{0.797(5)}$Fe$_{0.205(5)}$)OH(Fe$_{0.915(4)}$Li$_{0.085(4)}$)Se. However, true errors of the stoichiometric indices are certainly higher and rather about $\pm$10\,\%. Within this range, EDX measurements confirm the contents of iron, selenium and oxygen. Lithium was determined by ICP and hydrogen by elementary analysis to 0.8(3) wt-\% in general agreement with the expected 0.613 wt-\%.

\subsection*{Electrical transport and magnetic properties}

\begin{figure}[h]
\centering
\includegraphics[width=8cm]{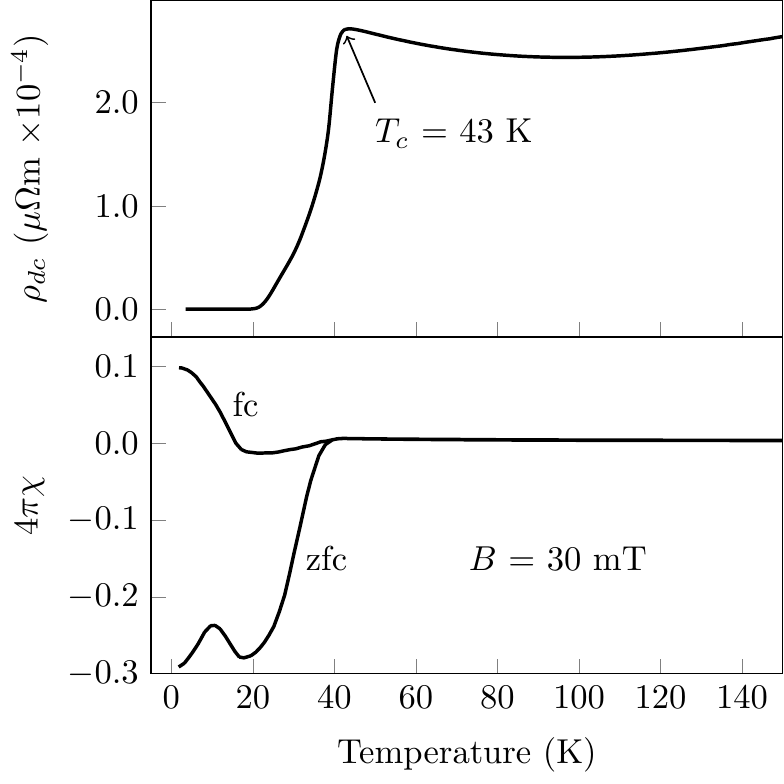}
\caption{\label{fig:lf+chi}Upper panel: $dc$-resistivity of the sample; lower panel: $dc$-magnetic susceptibility of \lfo.} 
\end{figure}

Fig.~\ref{fig:lf+chi} shows electrical transport and low-field magnetic susceptibility measurements of the \lfo\ sample. The resistivity is relatively high at 300\,K and weakly temperature dependent until it drops abruptly at 43\,K. Zero resistivity is reached below  25\,K. The superconducting transition is confirmed by the magnetic susceptibility which becomes strongly diamagnetic below 40\,K in a 30 mT field. However, the low temperature susceptibility behaves quite unusual. After zero field cooling (zfc, Fig.~\ref{fig:lf+chi}) the value starts strongly negative according to the shielding effect, and firstly increases with temperature until a maximum is reached at 10\,K, then decreases again until 18\,K, and finally increases steeply to zero as the temperature approaches $T_c$. In field cooled mode (fc, Fig.~\ref{fig:lf+chi}), the susceptibility becomes slightly negative below 40\,K due to the Meissner-Ochsenfeld effect, but increases again to positive values at lower temperatures. Remarkably, the diamagnetism of the superconductor competes with strong paramagnetism which emerges below 18\,K. The latter dominates in the fc mode, where the diamagnetic contribution due to the Meissner effect is weak. Thus actually no Meissner phase exists at the lowest temperatures.  Reversely in zfc mode the diamagnetic shielding is much stronger than the paramagnetic contribution. Note that the resistivity remains zero at low temperatures, which means that the emerging paramagnetic field is not strong enough to destroy the superconductivity. 

\begin{figure}[h]
\centering
\includegraphics[width=9cm]{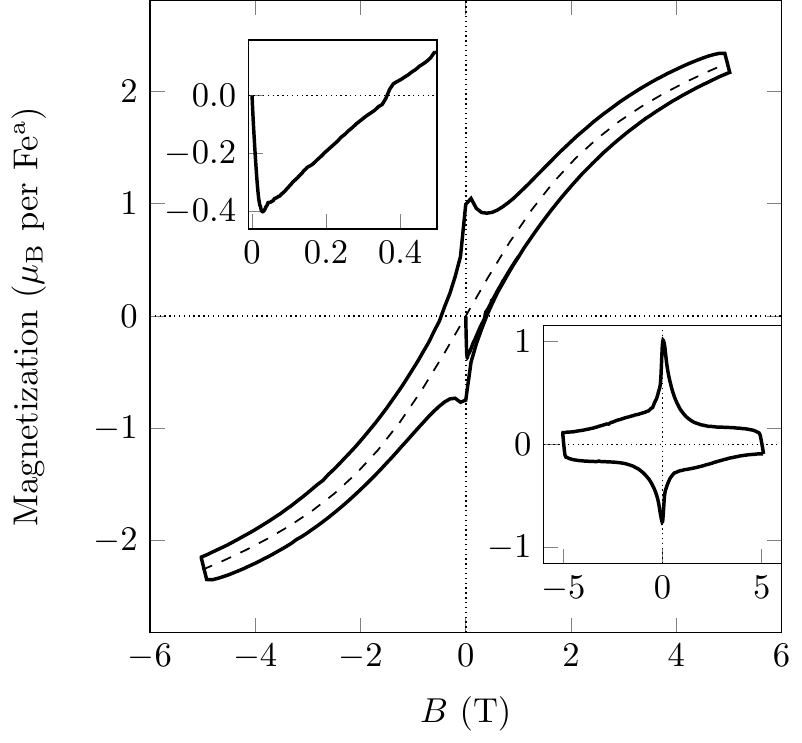}
\caption{\label{fig:mag}Isothermal magnetization at 1.8 K. Left inset: magnification of the low field part showing the initial curve. Right inset: magnetization after subtraction of the approximate ferromagnetic contribution (dashed line in the main plot). } 
\end{figure}

The magnetic susceptibility experiment suggests that superconductivity coexists with ferromagnetic ordering which emerges near 10\,K, well below the critical temperature of 43\,K. Fig.~\ref{fig:mag} shows the isothermal magnetization measured at 1.8\,K. The typical ferromagnetic hysteresis is superimposed by the magnetization known for hard type-II superconductors \cite{Buckel-2004, Bean-1962}. This becomes obvious if the approximate ferromagnetic contribution (dashed line in Fig.~\ref{fig:mag}) is subtracted. The resulting curve (inset in Fig.~\ref{fig:mag}) is typical for a superconductor which is partially penetrated by magnetic flux lines. Some flux becomes trapped due to vortex pinning, therefore we detect non-zero magnetization even at zero external field. The upper critical field of the superconductors is not reached at 5\,T, where the magnetization makes a typical jump because the sign of the field change reverses, and by this the directions of the shielding currents are also reversed. 

Unlike to the Chevrel phases or ErRh$_4$B$_4$ where the ferromagnetism destroys superconductivity, we observe the rare case where both phenomena can  co-exist because the ferromagnetic dipole field is smaller than the upper critical field of the superconductor. Given that the magnetization emerges inside the sample due to ferromagnetic ordering and  not by an external field, our material is in a so-called spontaneous vortex state. This is a new state of matter, where both orders coexist because the combined state has a lower free energy \cite{Greenside-1981}. Similar behaviour has been suggested in EuFe$_2$(As$_{1-x}$P$_x$)$_2$ where ferromagnetic ordering of Eu$^{2+}$ (4$f^7$) coexists with superconductivity \cite{Cao-2011,Nandi-2014}. In our case the ferromagnetism originates from the iron atoms in the (Li$_{1-x}$Fe$_x$)OH layer (\textit{vide infra}), thus \lfo\ is to our best knowledge the first example where superconductivity coexists with 3$d$-ferromagnetism, and moreover at the highest temperatures so far. 

\subsection*{Local probes} 

\begin{figure}[h]
\centering
\includegraphics[width=9cm]{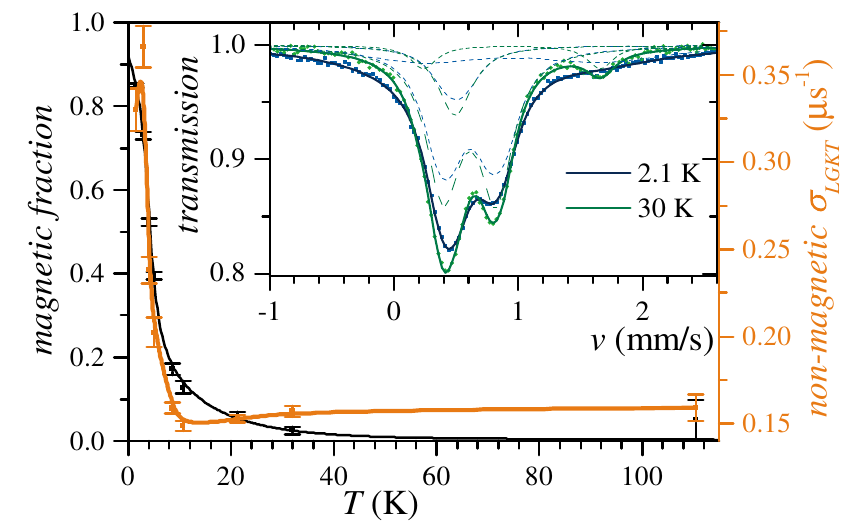}
\caption{\label{local}Magnetic volume fraction obtained by Zero field $\mu$SR shows that almost the whole sample is ruled by ferromagnetism at low temperatures. The increase of the static relaxation rate $\sigma_{LGKT}$ is due to the ferromagnetic stray field in the non magnetic sample fraction.  $^{57}$Fe M\"ossbauer spectra detect two iron sites and prove that magnetism arises from iron in the hydroxide layer, whereas the FeSe layer only senses stray fields.}
\end{figure}

The $^{57}$Fe M\"ossbauer spectrum (insert in Fig.~\ref{local}) consists of two doublets with an intensity ratio 0.9:0.1 in agreement with \feb\ in the FeSe and \fea\ in the hydroxide layers. The isomer shift  of $\approx 0.8$\,mm/s for \fea\ is typical for Fe$^{2+}$ in a $S$ = 4/2 state. The \fea\ doublet considerably broadens below $T_{\rm fm}\approx$\,10\,K which can be described by a hyperfine field of 3\,T at 2.1\,K. A small stray field of 0.4~T arising most propably from these ordered moments broadens the doublet at the \feb\ site in the FeSe layer. The remanence  of the internal fields proves the ferromagnetism. A third small subspectrum supports the asymmetry of the spectrum and suggests \feb\ sites in the FeSe layer with Li neighbors at the iron position.

%muSR
Zero field $\mu$SR data confirm the homogeneity of the sample as well as  ferromagnetic ordering below $T_{\rm fm}\approx$\,10\,K. The magnetism develops gradually  while the whole sample is ferromagnetic at 1.5\,K. A nonmagnetic fraction senses enhanced damping due to static fields from the magnetically ordered layer below 10\,K. Reducing the field cooled flux of 200\,G to 170\,G we can successfully prove bulk superconductivity by pinning nearly 40\% of the flux at 15\,K, confirmed by transverse field (TF) data at 200\,G. Cooling the sample from 40\,K down to 15\,K a considerable damping of the precession signal in more than 40\% of the sample is induced most probably due to flux line lattice formation and not ferromagnetism. However both TF and pinning experiments at 1.5\,K indicated, that superconducting volume fractions are retrieved by ferromagnetism.

\begin{figure}[h]
\centering
\includegraphics[width=9cm]{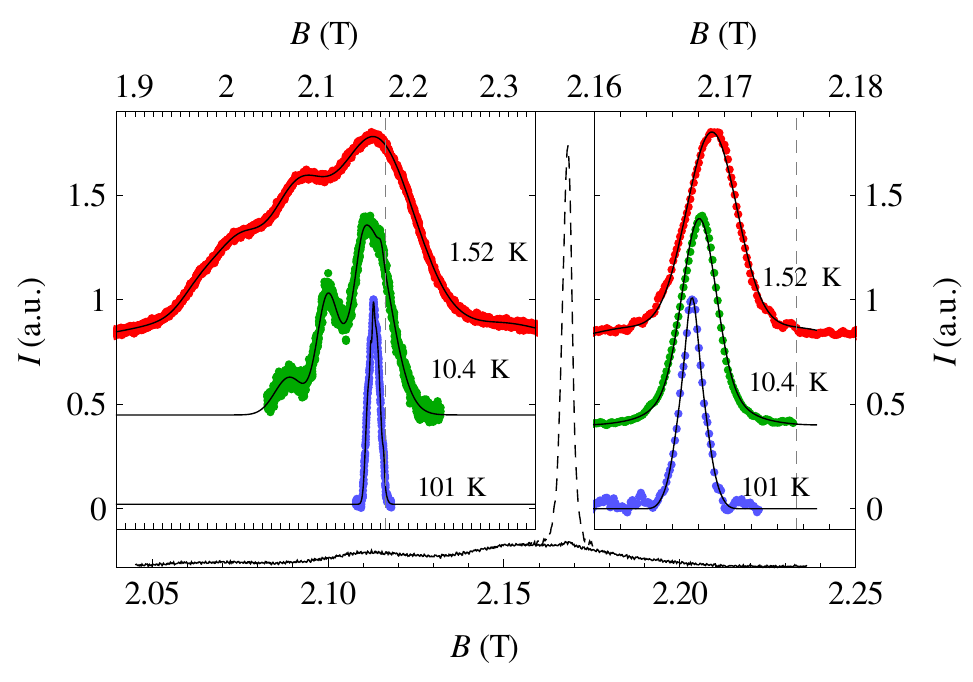}
\caption{\label{nmr}$^7$Li NMR spectra of \lfo. Outer graph: NMR spectra at $10~\mathrm{K}$ with a short repetition time (solid line) and with a long repetition time (dashed line). Left inset: temperature dependency of the broad fraction. Right inset: temperature dependency of the narrow fraction.}
\end{figure}

\bigskip 

High resolution \cite{Rees-2013} $^7$Li NMR spectra taken at $36~\mathrm{MHz}$  show two signal fractions originating from two Li sites (Fig.~\ref{nmr}), while the respective characteristic $T_1$ relaxation times are spread over $\approx 3$ orders of magnitude. The main fraction is broad and relaxes very fast. At low temperatures, the spectrum splits in mainly 3 broadened peaks. In contrast to that, the other fraction is a narrow line and relaxes very slowly. The spectrum shifts and its shape broadens slightly at low temperatures. Comparing the intensities of both spectral fractions, we assign the broad spectrum to lithium in the (Li$_{1-x}$Fe$_x$)OH layer,which is in line with the M\"ossbauer results. The vicinity to the magnetic \fea\  leads to broadening and a large shift. Because \fea\  is statistically distributed, different Li-surroundings produce a complex peak structure at low temperatures. The narrow spectrum is assigned to the Li at the \feb-sites in the nonmagnetic FeSe layer. The small shift and broadening at low temperatures is due to stray fields and vanishing Pauli magnetism in the superconducting phase.

\subsection*{Electronic structure calculations} 

DFT band structure calculations with an ordered model of [(Li$_{0.8}$Fe$^{\rm a}_{0.2}$)OH]Fe$^{\rm b}$Se according to [Li$_4$Fe$^{\rm a}$(OH)$_5$(Fe$^{\rm b}$Se)$_5$] were carried out. First the atomic coordinates of a $\sqrt{5}a\times\sqrt{5}a$ superstructure were allowed to relax, then we tried different ordering patterns. No magnetic ground state with non-zero moments at the Fe$^{\rm b}$-site in the FeSe-layer could be obtained. On the other hand, ferromagnetic ordering of the moments at the Fe$^{\rm a}$-site in the hydroxide layer lowered the total energy by 41 kJ/mol with a magnetic moment of 3.5\,$\mu_{\rm B}$ per Fe$^{\rm a}$. Antiferromagnetic ordering resulted in the same stabilization, thus our model cannot distinct between ordering patterns, but it definitely shows that magnetism emerges from iron in the hydroxide layer.   

\begin{figure}[h]
\centering
\includegraphics[width=9cm]{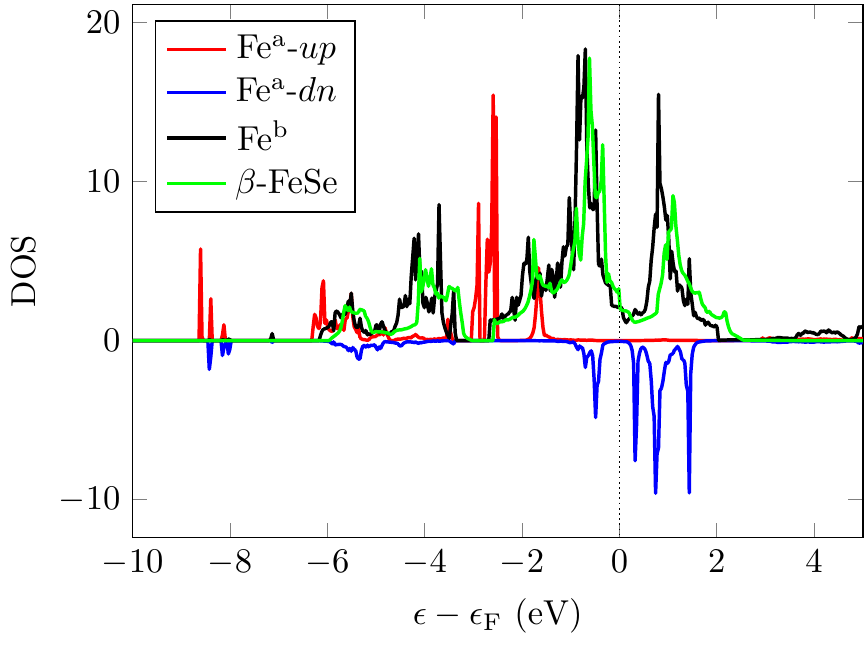}
\caption{\label{fig:dos}Electronic density of states (DOS) contributions of the iron atoms. Red/blue: Magnetic Fe$^{\rm a}$ atoms in the hydroxide layer; black: nonmagnetic Fe$^{\rm b}$ atoms in the FeSe layer; green: Fe-atoms in binary $\beta$-FeSe for comparison.}
\end{figure}

Fig.~\ref{fig:dos} shows the contributions of the different iron atoms to the electronic density of states (DOS). The magnetic exchange splitting of the Fe$^{\rm a}$ states is clearly discernible, while the states of the nonmagnetic Fe$^{\rm b}$ sites remain almost exactly as in binary $\beta$-FeSe (green line in Fig.~\ref{fig:dos}). Moreover, the Fermi-level is located just in a gap of the magnetic Fe$^{\rm a}$ states. This means that the electronic systems of the individual layers interact very weakly, and that the typical Fermi-surface topology known from other iron based superconductors \cite{Sunagawa-2014} is not disturbed by the presence of the hydroxide layer. Nevertheless the latter acts as an electron reservoir. Formally 0.2 electrons are transferred from the hydroxide to the selenide layer according to [(Li$_{0.8}$Fe$^{2+}_{0.2}$)OH]$^{0.2+}$(FeSe)$^{0.2-}$. This is also evident from the small shift of the Fe$^{\rm b}$ states (black line in Fig.~\ref{fig:dos}) to lower energies relatively to $\beta$-FeSe. We suggest that this electron doping of the FeSe layer is mainly responsible for the enormous increase of $T_c$ in our compound (43\,K) in comparison to $\beta$-FeSe (8\,K). Similar electron transfers of $\approx$\,0.2 $e^-$/FeSe have recently been reported for other intercalated iron selenides, among them Li$_x$(NH$_2$)$_y$(NH$_3$)$_{1-y}$Fe$_2$Se$_2$ ($T_c$ = 43\,K) \cite{Burrard-Lucas-2013,Sedlmaier-2013}, K$_x$Fe$_2$Se$_2$ ($T_c$ = 32\,K) \cite{Carr-2014}, Na$_x$Fe$_2$Se$_2$ ($T_c \approx$ 46\,K) \cite{Ying-2012, Ying-2013}, and Li$_x$(C$_2$H$_8$N$_2$)$_y$Fe$_{2-z}$Se$_2$ ($T_c \approx$ 45\,K) \cite{Hatakeda-2013}.  

\bigskip 
                            
\section*{Conclusion}

Superconductivity  below $T_c$ = 43\,K coexists with ferromagnetism  below $T_{\rm fm} \approx$  10\,K in \lfo\ synthesized under hydrothermal conditions. The layered crystal structure consists of ferromagnetic (Li$_{1-x}$Fe$_x$)OH and superconducting Fe$_{1-x}$Li$_x$Se layers each with $anti$-PbO-type structures. Both physical phenomena are spatially separated, but the internal dipole field of the ferromagnet acts on the superconductor, which suggests the existence of a special state of matter called spontaneous vortex phase. The local probes $^{57}$Fe-M\"ossbauer spectroscopy, $^7$Li-NMR and $\mu$SR consistently support this conclusion. This rare phenomenon was so far confined to $f$-shell magnetism, while in our compound superconductivity coexists with 3$d$-ferromagnetism for the first time, and moreover at the highest temperatures so far. In contrast to the chemically inert $f$-shells, 3$d$-magnetism is much more susceptible to the chemical environment, which opens new avenues for chemical modifications that can now directly couple to the magnetic and superconducting properties, thus allowing broader studies of such coexistence phenomena in the future.           

\section*{Acknowledgement}
The authors thank the German Research Foundation (DFG) for financial support of this work within SPP1458 and GRK1621.  

\section*{Methods and materials}

Polycrystalline samples of \lfo\ were synthesized under hydrothermal conditions using a modified procedure given in \cite{Lu-2014}.  0.0851\,g iron  metal (99.9\,\%), 0.5\,g Selenourea (99\,\%) and 3\,g LiOH$\cdot$H$_2$O were mixed with 10 mL distilled water. The starting mixtures were tightly sealed in a teflon-lined steel autoclave (50 mL) and heated at 150$^{\circ}$\,C for 8 days. The obtained shiny lamellar precipitates were separated by centrifugation, and washed several times with distilled water and ethanol. Afterwards, the polycrystalline products were dried at room temperature under dynamic vacuum and stored at $-25^{\circ}$\,C under argon atmosphere.

X-ray powder diffraction was carried out using a Huber G670 diffractometer with Cu-K$\alpha_1$ radiation ($\lambda$ = 154.05 pm) and Ge-111 monochromator. Structural parameters were obtained by Rietveld refinement using the software package TOPAS \cite{Topas}. Single crystal analysis was performed on a  of Bruker D8-Quest diffractometer (Mo-K$\alpha_1$,  $\lambda$ = 71.069 pm, graphite monochromator). The structure was solved and refined with the Jana2006 program package \cite{Jana}. 
Chemical compositions were additionally determined by energy dispersive X-ray  analysis (EDX) as well as by chemical methods using ICP-AAS and elemental analysis.  Magnetic properties were examined with a Quantum Design MPMS-XL5 SQUID magnetometer, whereas temperature dependent resistivity measurements were carried out using a standard four-probe method.

$^{57}$Fe-M\"ossbauer spectroscopy was performed with a standard Wissel setup in transmission geometry using a Co/Rh source with an experimental line width $\omega_{exp}=$~0.13\,mm/s. $\mu$SR experiments were carried out with the GPS spectrometer at the $\pi$M3.2 beamline of the Swiss Muon Source at the Paul Scherrer Institut in Villingen, Switzerland. $^7$Li Nuclear magnetic resonance spectra are taken at several temperatures using the Fourier-transformation field-sweep method. 

Electronic structure calculations were performed using the Vienna ab initio simulation package (VASP) \cite{Kresse-1994,Kresse-1996}, which is based on density functional theory (DFT) and plane wave basis sets. Projector-augmented waves (PAW) \cite{Bloechl-1994} were used and contributions of correlation and exchange were treated in the generalized-gradient approximation (GGA) as described by Perdew, Burke and Ernzerhof \cite{Perdew-1996}. 

\pagebreak

\bibliographystyle{unsrt}
%\bibliography{LFOFS}

%\usepackage[english]{babel}
%\usepackage[latin1]{inputenc}
%\usepackage[T1]{fontenc}
%\usepackage{amsmath}
%\usepackage{graphicx}
%\usepackage{tabularx}

\section*{Supporting Information}

\setcounter{figure}{0} \renewcommand{\thefigure}{S\arabic{figure}}

\setcounter{table}{0} \renewcommand{\thetable}{S\arabic{table}}

\subsection*{M\"ossbauer Spectroscopy}
$^{57}$Fe-M\"ossbauer spectroscopy was performed with a standard Wissel setup in transmission geometry using a Co/Rh source with an experimental line width (HWHM) $\omega_{exp}=$~0.13\,mm/s. Spectra were recorded in a warming series and analysed using a static Hamiltonian approach
	\[ H_s = \frac{eQ_{zz}V_{zz}}{4I(2I-1)}\left[(3I_z^2-I^2)\right]  -g_I\mu_NB\left(\frac{I_+e^{-i\Phi}+I_-e^{+i\Phi}}{2}\sin\Theta+I_z\cos\Theta\right)\]
setting the field gradient asymmetry to zero. The a- and b-site alone do not sufficiently reproduce the asymmetry of the main b-doublet. Rather appear 20\% of the b-site to have different hyperfine parameters, thus enlarging the left peak. These 20\% perfectly coincide with a 7\% Fe/Li mixing, which leads to around 20\% disturbed iron surroundings considering randomly distributed Li in a binomial distribution. However a full distribution of field gradients would be the correct physical interpretation, especially with regard on the bigger line width of the b-doublet compared to the a-doublet in the paramagnetic state. A different interpretation of the missing asymmetry is a texture effect, which could account for a more pronounced left peak in both subspectra, caused by the flaky shape the crystallites. The absolute value of quadrupole splitting of 0.41\,mm/s is relatively large compared to the 0.29\,mm/s for the pure FeSe \cite{blachowski2010mossbauer}, which might be easily explained by the additional interlayer and the orientation of the principal axis in c-direction. The quadrupole splitting of 1.41\,mm/s of the a-site was fitted globally.

\begin{figure}[h]
\centering
\includegraphics[width=0.4\paperwidth]{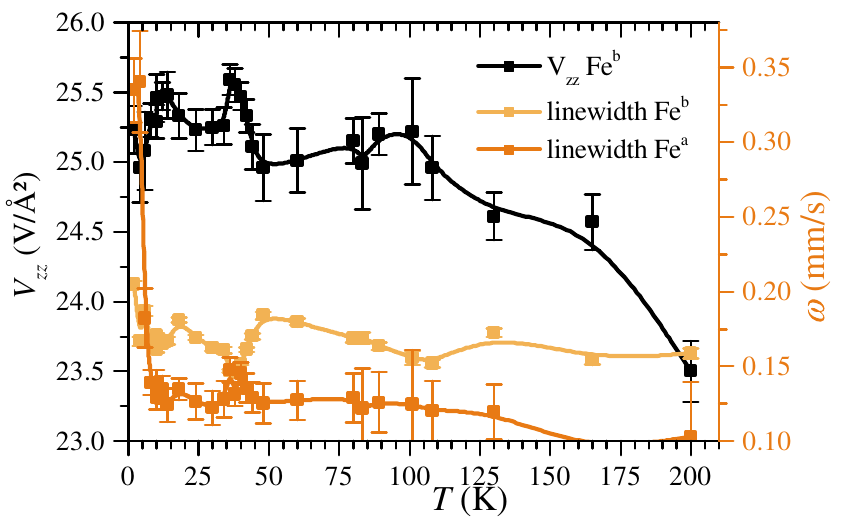}
\caption{The changes of the b-doublet shape indicate the superconducting transition at 40\,K. While the ferromagnetism at the Fe\textsuperscript{a} leads to a huge line broadening and a hyperfine field of $\approx$~3~T, the doublet of the Fe\textsuperscript{b} in the FeSe layer only slightly broadens}\label{figMBS-HWHMandVzz}
\end{figure}

The Fe\textsuperscript{b} site does not contain magnetic hyperfine field contribution whereas Fe\textsuperscript{a} needs a real magnetic splitting to sufficiently describe the subspectrum in the ferromagnetic regime additionally to the huge line broadening. The temperature dependencies of the principal component $V_{zz}$ of the field gradient and the line widths $\omega$ are shown in Fig. \ref{figMBS-HWHMandVzz}. They indicate experimentally the transition temperatures of the superconductivity and the ferromagnetism, although these parameters are correlated, i.g. they partially compensate each other.

Both, the quadratic Doppler effect and Debye Waller factor do not show any anomaly and can roughly be fitted simultaneously in a Debye approximation with a Debye temperature of 192\,K.

\pagebreak

\subsection*{$\mu$SR}

\begin{figure}[h]
\centering
\includegraphics[width=0.4\paperwidth]{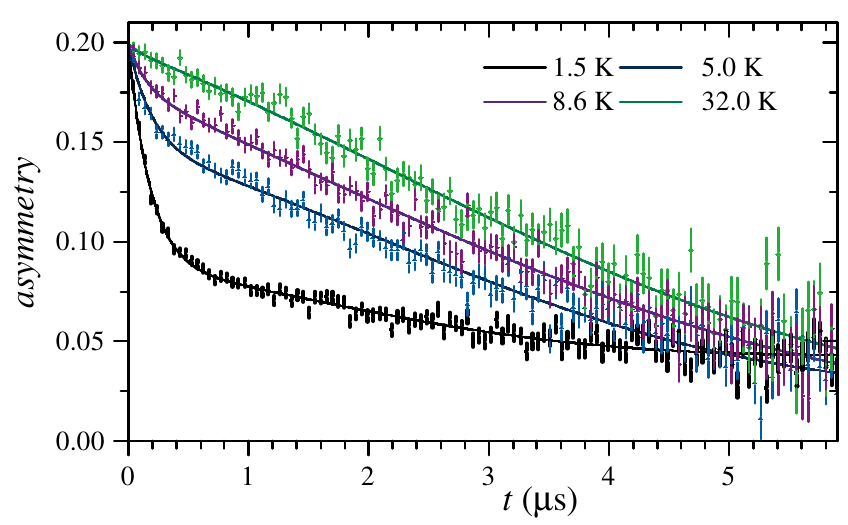}
\caption{The virtual constant 1/3-tail of the 1.5\,K spectrum reveals almost 100\% magnetic fraction. The nonmagnetic fraction is considerably influenced by stray fields.}\label{figmuSR-ZFspectr}
\end{figure}

$\mu$SR experiments were carried out at the GPS spectrometer at the $\pi$M3.2 beamline of the Swiss Muon Source at the Paul Scherrer Institut. Zero field (ZF) time spectra (Fig. \ref{figmuSR-ZFspectr}) reveal a almost 100\% magnetic fraction at base temperature with regard on the 1/3-tail of the lowest temperature spectrum.

\begin{figure}[h]
\centering
\includegraphics[width=0.4\paperwidth]{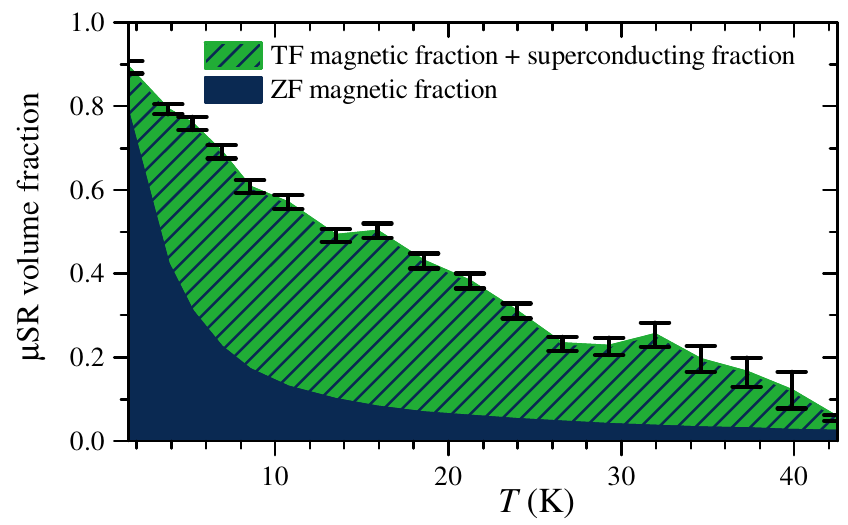}
\caption{Transverse field data was analyzed using a ferromagnetic fraction fixed to ZF data. The residual signal can be described by an Gaussian damping, which might include aside superconducting origin also ferromagnetic contributions. However the pinning experiment (Fig. \ref{figpinning}) at 15\,K proves this fraction indeed to be of superconducting origin.}\label{figmuSR-200Gfrac}
\end{figure}

\begin{figure}[h]
\centering
\includegraphics[width=0.48\textwidth]{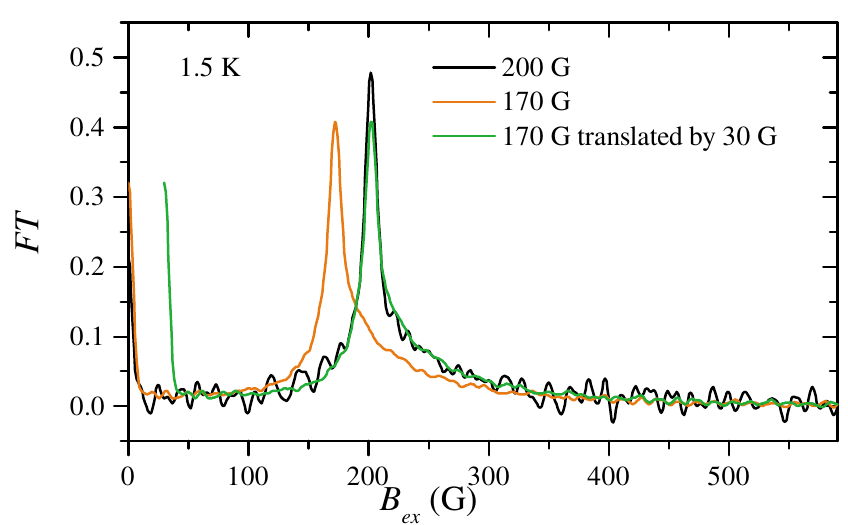}
\includegraphics[width=0.48\textwidth]{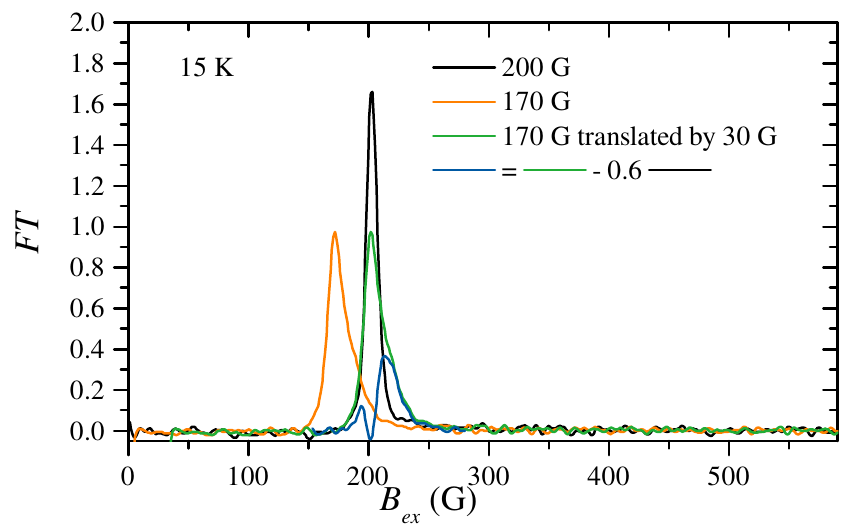}
\caption{The Fourier transformation $FT$ of the time spectra of the $\mu$SR pinning experiments prove bulk superconductivity above the Curie temperature at 15\,K (right). In contrast at 1.5\,K (left) pinning is absent, thus the complete suppression of superconductivity by ferromagnetism is probable.}\label{figpinning}
\end{figure}

The asymmetry $A(t)$ in the whole temperature range was fitted using the following two fraction model.
\[ \frac{A_{ZF}(t)}{A_0} = f_{fm}\cdot\left(\frac{2}{3}e^{\lambda_Tt}+\frac{1}{3}e^{\lambda_Lt}\right)+(1-f_{fm})\cdot\left(\frac{2}{3}(1-\sigma_{LGKT}^2t^2-\lambda t)e^{-\sigma_{LGKT}^2t^2/2-\lambda t}+\frac{1}{3}\right) \]
The parameter $\lambda$ was fixed to the 100\,K value, whereas $\sigma_{LGKT}$ represents the damping due to static stray fields, which was presented in the main text. The transverse damping rate of the magnetic fraction ranges between 6 and 10~$\mu s^{-1}$, peaking at 5\,K, whereas the longitudinal rate $\lambda_L$ stays almost close to zero except some barely significant increase at 10\,K.

Transverse field (TF) experiments were done to deduce superconducting volume fractions and order parameters. Usually the superconducting order parameter is deduced from the Gaussian damping rate of the precession signal. As a matter of fact the signal at 1.5\,K is damped exponentially due to ferromagnetism, but there is some considerable Gaussian fraction at temperatures between 10\,K and 40\,K. The change of damping behavior can be displayed by a stretched exponential fit, however this is no physically reasonable. A three fraction (ferromagnetic, superconducting, paramagnetic) fit only works with fixed ferromagnetic fractions $f_{fm}$ from the zero field analysis:

\begin{align}
	\frac{A_{TF}(t)}{A_0} &= f_{fm,ZF}\cdot\cos(\gamma (B_{ex}+dB_{fm})t+\phi_0)e^{-\lambda_{FM}t}\\
	&+f_{SC}\cdot\cos(\gamma (B_{ex}+dB_{SC})t+\phi_0)e^{-\sigma{SC}^2t^2/2}\\
	&+(1-f_{fm}-f_{SC})\cdot\cos(\gamma B_{ex}t+\phi_0)e^{-\lambda_{pm}t}
\end{align}

The ferromagnetic damping $\lambda_{FM}$ stays almost constant at $\approx$7~$\mu s^{-1}$, the paramagnetic damping $\lambda_{pm}$ was fixed to the 45\,K value. The damping rate $\sigma_{SC}$ , which is assigned mainly to superconductivity but might also include additional ferromagnetic damping, increases from 42\,K to 27\,K from zero to 0.6~$\mu s^{-1}$. As the ferromagnetic and superconducting fractions are not clearly distinguishable by these TF measurements we resign to interpret the superconducting order parameter. In Fig. \ref{figmuSR-200Gfrac} the temperature dependence of the volume fractions is shown. A superconducting volume fraction of 40\% maximum can be deduced from this 200\,G-TF data. Additional 700\,G measurements were even worse to analyse.

As the situation from TF measurements was not clear we decided to perform pinning experiments. Field cooling the sample at 200\,G starting from temperatures above 45\,K should lead to a well established flux line lattice in the superconducting parts of the sample, which in case of sufficient amount and strength of pinning centers, should be kept in the superconducting state when the field is reduced. This indeed is the case for $\approx$~40\% of the signal at 15\,K (Fig. \ref{figpinning}), in agreement with the TF data. In contrast, the pinning fails at 1.5\,K. Most probably, the superconductivity is then suppressed by the ferromagnetism.

\pagebreak

\subsection*{Crystallographic data}

\begin{table}[h]
\caption{Crystallographic data of (Li$_{0.795(5)}$Fe$_{0.205(5)}$)OH(Fe$_{0.915(4)}$Li$_{0.085(4)}$)Se}
\begin{tabular}{ll} 
\hline
Formula        &              (Li$_{0.795(5)}$Fe$_{0.205(5)}$)OH(Fe$_{0.915(4)}$Li$_{0.085(4)}$)Se   \\
Formula weight  (g mol$^{-1}$)&                                   164.6  \\
Crystal System &                                         Tetragonal   \\
Space group  &                                 $P4/nmm$ O1  (No.129)   \\
$a$, $c$ (pm)      &          380.38(1),  922.10(6)   \\
$V$ (nm$^3$)        &                                          0.13342(1)   \\
$Z$              &                                                  2   \\
$d_{calc}$ (gcm$^3$) &                                           4.097 \\  
$\mu$ (Mo-K$\alpha$) (mm$^{-1}$)   &                                         19.6  \\
Crystal Size ($\mu$m$^3$)    &                         50 $\times$ 30 $\times$ 5     \\
Temperature (K)       &                                         293  \\
Radiation (pm)   &                          Mo-K$\alpha$    $\lambda$ = 71.073  \\
$\theta$ range (deg)     &                                2.2 - 33.1  \\
$hkl$ range                  &               $\pm 5$;  $\pm 5$; $-$10$\rightarrow$+14  \\
Tot., Uniq. Data, $R_{\rm int}$  &                    1611,    189,  0.027  \\
$N_{Refl}$, $N_{Par}$                        &                       189,   15  \\
$R$1, $w$R$_2$, $S$            &           0.0161, 0.0432, 1.35 \\                                
$\Delta\rho_{\rm min}$, $\Delta\rho_{\rm max}$,  ($e$\AA$^{-3}$) &                  $-$0.58, 0.43 \\
\hline
\end{tabular}

\begin{tabular}{l}
Atomic positions and equivalent displacement parameters
\end{tabular}

\begin{tabular}{lllllll}
Atom &          $Wyck.$ & $x$ & $y$ & $z$ & occ. & $U_{\rm eq}$ \\ 
Li1 &           $4f$    &  0 & 0  & 0.043(6) & 0.795(5) & 0.031(3) \\
Fe$^{\rm a}$ &  $2a$    &  0 & 0  & 0        & 0.205(5) & 0.031(3)\\ 
O            &  $2c$    &  0 & 1/2 & 0.0716(4) & 1.0 & 0.0194(7)\\
H            &  $2c$    &  0 & 1/2 & 0.152(6) & 1.0 & 0.02\\
Fe$^{\rm b}$ &  $2b$    &  0 & 0   & 0.5      & 0.915(4) & 0.0151(1)\\
Li2          &  $2b$    &  0 & 0   & 0.5      & 0.085(4) & 0.0151(1) \\
Se & $2c$ & 1/2  & 0 & 0.33874(4) & 1.0 & 0.0153(1) \\
\hline
\end{tabular}   

\begin{tabular}{l}
Selected bond lengths (pm) and angles (deg)
\end{tabular}

\begin{tabular}{lllllll}
Li-O & 192.0(7)$\times$2 & 218(2)$\times$2 & O-Li-O & 164(1) & 122(1) & 93.8(1)$\times$4\\  
Fe$^{\rm a}$-O & 201.3(1)$\times$4 & ~ & O-Fe$^{\rm a}$-O & 141.7(1)$\times$2 & 96.2(1)$\times$4  \\ 
Fe$^{\rm b}$-Se & 241.42(2)$\times$4 & ~ & Se-Fe$^{\rm b}$-Se & 103.96(1)$\times$2 & 112.30(1)$\times$4 \\
\hline
\end{tabular}

\end{table}

\bibliographystyle{plain}
%\bibliography{bibliography}

\end{document}